\documentclass[conference]{IEEEtran}
\IEEEoverridecommandlockouts
\usepackage{cite}
\usepackage{amsmath,amssymb,amsfonts}
\usepackage{graphicx}
\usepackage{textcomp}
\usepackage{amsmath} 
\usepackage{amsfonts} 
\usepackage{amssymb} 
\usepackage{algorithm}
 \usepackage{algpseudocode}

\usepackage{booktabs} 
\usepackage{multicol,multirow} 
\usepackage[table]{xcolor} 
\usepackage{colortbl}
\usepackage{xspace}
\usepackage{threeparttable}
\usepackage{subcaption}

\def\BibTeX{{\rm B\kern-.05em{\sc i\kern-.025em b}\kern-.08em
    T\kern-.1667em\lower.7ex\hbox{E}\kern-.125emX}}
\begin{document}

\title{Strategic Pseudo-Goal Perturbation for Deadlock-Free Multi-Agent Navigation in Social Mini-Games \\
}

\author{\IEEEauthorblockN{Abhishek Jha*}
\IEEEauthorblockA{\textit{Department of Mechanical Engineering} \\
\textit{Delhi Technological University}\\
New Delhi, India \\
aplusj247@gmail.com}
*Corresponding author
~\\
\and
\IEEEauthorblockN{Tanishq Gupta}
\IEEEauthorblockA{\textit{Department of Mechanical Engineering} \\
\textit{Delhi Technological University}\\
New Delhi, India \\
gtanishq9@gmail.com}

~\\
\and
\IEEEauthorblockN{Sumit Singh Rawat}
\IEEEauthorblockA{\textit{Department of Mechanical Engineering} \\
\textit{Delhi Technological University}\\
New Delhi, India \\
singhrawat.sumit@gmail.com}
~\\
\and
\IEEEauthorblockN{Girish Kumar}
\IEEEauthorblockA{\textit{Department of Mechanical Engineering} \\
\textit{Delhi Technological University}\\
New Delhi, India \\
girishkumar@dce.ac.in}

}

\maketitle

\begin{abstract}
This work introduces a Strategic Pseudo-Goal Perturbation (SPGP) technique, a novel approach to resolve deadlock situations in multi-agent navigation scenarios. Leveraging the robust framework of Safety Barrier Certificates, our method integrates a strategic perturbation mechanism that guides agents through social mini-games where deadlock and collision occur frequently. The method adopts a strategic calculation process where agents, upon encountering a deadlock select a pseudo goal within a predefined radius around the current position to resolve the deadlock among agents. The calculation is based on controlled strategic algorithm, ensuring that deviation towards pseudo-goal is both purposeful and effective in resolution of deadlock. Once the agent reaches the pseudo goal, it resumes the path towards the original goal, thereby enhancing navigational efficiency and safety. Experimental results demonstrates SPGP's efficacy in reducing deadlock instances and improving overall system throughput in variety of multi-agent navigation scenarios.
\end{abstract}


\section{Introduction}
In the domain of control and robotics, multi-agent navigation poses significant challenges, particularly when multiple autonomous agents operate within the same environment. These challenges are compounded in scenarios where agents must navigate social spaces and collisions among agents deemed to happen which are referred to as "social mini-games"—such as doorways, intersections, and L-corners. These environments necessitate not only collision avoidance but also efficient coordination among agents to prevent deadlock situations, where agents halt progress due to mutual obstruction \cite{b1,b2, b3, b4, b5, b6}.

Collision avoidance in multi-agent systems is a well-studied area, with numerous algorithms developed to enable autonomous agents to navigate without incidents \cite{b5, b6, b7}. But, the challenges of social mini-games introduces complex interaction dynamics which is not fully solved by the existing collision avoidance techniques. In these scenarios, agents often encounter deadlock due to the limited space for the agents to navigate and the symmetry found in the environment  \cite{b8, b9}. Such deadlocks significantly restricts the flow of movement, leading to low efficiencies and performance. 

\begin{figure}[htbp]
    \centering
    \includegraphics[width=0.45\textwidth]{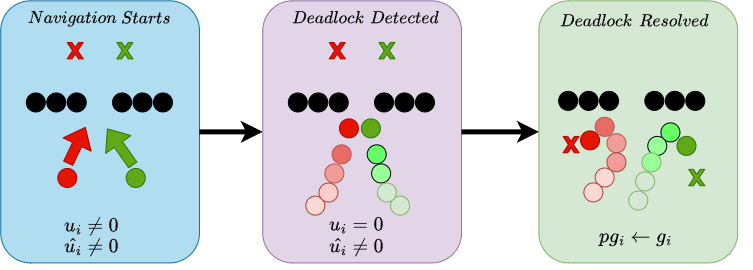}
    \caption{Overview of Strategic Pseudo-Goal Perturbation}
    \label{fig:overview}
\end{figure}

Our method, Strategic Pseudo-Goal Perturbation (SPGP), addresses the challenge of deadlock in multi-agent navigation within social mini-games. SPGP innovates by combining Safety Barrier Certificates with a mechanism that, upon detecting a potential deadlock, strategically perturbs agents towards pseudo goals. These pseudo goals are generated based on a strategic algorithm that enforces efficient deviation from the agent's original path, facilitating the resolution of the deadlock. Once the agent reaches its pseudo goal, it recalculates its path towards the original goal, thus minimizing errors and increasing the efficiencies in this process.
By concentrating on strategic perturbation rather than other techniques, SPGP offers solution to resolve deadlock in complex social navigation scenarios. This perturbation technique not only the efficiency of multi-agent navigation but also contributes to the improvement of autonomous agent coordination in constrained environments.
Fig.~\ref{fig:overview} depicts the overview of SPGP method where navigation starts then deadlock ios detected and finally deadlock is resolved using the strategic perturbation.

This study lays contributions to the field of multi-agent navigation and deadlock avoidance in social mini-game scenarios. The key contributions are as follows:
\begin{itemize}
    \item Introduced a Strategic Pseudo-Goal Perturbation (SPGP) method, a novel approach to generate deadlock-free situations in multi-agent navigation. This method strategically perturbs agents towards pseudo goals.
    
    \item Through iterative analysis, explored the variation of the perturbation radius and its effect on the effectiveness of deadlock resolution, providing insights into optimal perturbation strategies for varying environmental complexities and agent densities.
    
    \item Thoroughly tested the SPGP method across various complex social mini-games scenarios. These experiments depicts the flexibility and adaptability of our method, which shows its effectiveness in a wide range of common social navigation challenges.

    \end{itemize}

\section{Related Work}

\subsection{Safety Based Methods}

The ORCA framework\cite{b10}, along with its non-holonomic extension \cite{b11}, shows provable safety in multi-agent navigation through single-integrator systems. This method efficiently generates optimal collision-free velocities by using linear programming to navigate constraints represented as half-planes in velocity space. While originally designed for holonomic systems, the framework's also been extended for non-holonomic constraints, aiming to impose minimal deviation from preferred velocities. However, the construction of half-planes in ORCA can lead to deadlocks, marking its primary limitation \cite{b12}. Ensuring safety in systems with double-integrator dynamics is more challenging, with safety largely influenced by the system's planning frequency. For instance, the NH-TTC framework \cite{b13} employs gradient descent to optimize a cost function that combines position with a time-to-collision component,  which helps to prevent imminent collisions. NH-TTC's safety assurances improve as planning frequency increases towards infinity. Another set of methods uses the shortest path in the multi-agent network to reach the goal position in a complex environment \cite{b29, b30}. Similarly, the model predictive control (MPC) method \cite{b8} underscores safety as a function of both planning frequency and the extent of the time horizon. Further, Control Barrier Functions (CBFs) \cite{b15,b16} offer a method to design controllers ensuring safety by maintaining the forward invariance property in the set, ensuring that if an agent starts within a safety set, it remains within this set indefinitely, effectively preventing it to move away from the safe zone.

\subsection{Learning Based Methods}

The integration of conventional navigation techniques with machine learning represents a fast-evolving research space on learning-based motion planning. Among the various techniques, deep reinforcement learning (DRL) and trajectory prediction stand out for their success in multi-agent planning and navigation. DRL is applied to generate navigation policies through simulation, including for scenarios involving multiple robots in social mini-games. For instance, \cite{b17} describes a decentralized collision avoidance system for multi-robot scenarios using DRL and local sensory inputs. Another example is CADRL \cite{b18}, which optimizes social robot navigation by employing a sparse reward system, with extensions like LSTM-based action selection in response to the proximity of other robots \cite{b19}. On the other hand, algorithms incorporating trajectory prediction focus on forecasting future robot positions to navigate dynamically among moving obstacles effectively. 

Another avenue is Imitation Learning (IL) which is a machine learning strategy where an agent learns behaviors by mimicking expert demonstrations, typically faster than conventional optimization methods. These demonstrations often include state input pairs from expert policy executions. The main IL methodologies are Behavior Cloning (BC) \cite{b20,b21}, which uses supervised learning to directly acquire an imitative policy; Inverse Reinforcement Learning (IRL) \cite{b22}, which deduces a reward function from demonstrations to facilitate policy learning via Reinforcement Learning (RL) \cite{b23}; and approaches utilizing generative models \cite{b24}. Despite their efficiencies, IL approaches \cite{b25,b26} commonly face limitations in encoding state/safety and input constraints, presuming access to expert action data during demonstrations and struggling to adapt to scenarios beyond the expert demonstrations' distribution.

\subsection{Deadlock Resolution Methods}

Recent studies have explored structured perturbation strategies, such as the right-hand rule \cite{b27} and clockwise rotation \cite{b15}, to enhance performance beyond random perturbation methods, even providing formal optimality guarantees. However, these approaches are constrained by a preset ordering, which restricts their applicability, often to scenarios with no more than three agents.
 Another different category of deadlock resolution techniques adopts priority and scheduling mechanisms which are generally found in managing intersection in autonomous agents research \cite{b28}. Notable among these protocols are First Come First Served (FCFS), reservation, and auction systems. Specifically, this method \cite{b29} prioritizes agents by their sequence of arrival at intersections. While straightforward to apply, this method may result in extended waiting periods and increased congestion when numerous vehicles converge on an intersection at the same time.

\section{Preliminaries}

In this section, we lay the basis for our contributions by discussing the foundational concepts of Control Barrier Functions (CBF), Safety Barrier Certificates (SBC) and Deadlock conditions which form the basis for ensuring safety in multi-agent navigation.

\subsection{Control Barrier Functions}

Control Barrier Functions (CBFs) provides assurance of the forward invariance property of the safe set for the agents\cite{b15, b17}.
\begin{equation}
    \dot{x} = f(x) + g(x)u,
\end{equation}
where \( x \in \mathbb{R}^n \) represents the state and \( u \in \mathbb{R}^m \) represents the control input, a set \( \mathcal{C} \subseteq \mathbb{R}^n \) is said to be forward invariant if, for the initial condition \( x(0) \in \mathcal{C} \), we have \( x(t) \in \mathcal{C} \) for all \( t \geq 0 \). The set \( \mathcal{C} \) is defined as the zero-superlevel set of a continuously differentiable function \( h: \mathbb{R}^n \rightarrow \mathbb{R} \), such that

\begin{equation}
    \mathcal{C} = \{ x \in \mathbb{R}^n \mid h(x) \geq 0 \}.
\end{equation}
The time derivative of \( h(x) \) around the trajectories of the agent's system is given by
\begin{equation}
    \frac{dh(x)}{dt} = \frac{\partial h}{\partial x} \cdot \frac{dx}{dt} = \frac{\partial h}{\partial x} \cdot (f(x) + g(x)u),
\end{equation}
or, using the Lie derivative form,
\begin{equation}
    \frac{dh(x)}{dt} = L_f h(x) + L_g h(x)u.
\end{equation}

A function \( h(x) \) is a CBF for the set \( \mathcal{C} \) if there exists an extended class-\(\mathcal{K}\) function \( \kappa \) such that for all \( x \in \mathbb{R}^n \),
\begin{equation}
    \sup_{u \in \mathbb{R}^m} [L_f h(x) + L_g h(x)u + \kappa(h(x))] \geq 0,
\end{equation}
where \( L_f h(x) \) and \( L_g h(x) \) denote the Lie derivatives of \( h \) along \( f \) and \( g \), respectively. For \( h(x) \) to serve as a CBF, it must satisfy the inequality
\begin{equation}
    L_f h(x) + L_g h(x)u + \gamma h^3(x) \geq 0, \quad \forall x \in \mathbb{R}^n,
\end{equation}
for some extended class-\(\mathcal{K}\) function \( \gamma h^3(x) \), assuring the system resides  within the safe set \( \mathcal{C} \) for every future timesteps. where \(\gamma\) represents the controlling parameter and \(h(x)\) represents the barrier function.

\subsection{Safety Barrier Certificates}
Safety Barrier Certificates (SBCs) are designed for enforcing collision avoidance among various agents in a multi-agent system \cite{b15}. For agents $i$ and $j$, with dynamics modeled as double integrator, the relative velocity and position vectors are denoted as $\Delta p_{ij}$ and $\Delta v_{ij}$ respectively. To ensure a safe distance $D_s$ is maintained, the joint constraint for safety barrier of agents $i \neq j$ is expressed in equation 7:
\begin{equation}
    A_{ij}u \leq b_{ij},
\end{equation}
\label{eq:ab}
where $A_{ij}$ and $b_{ij}$ are defined as follows:

\begin{itemize}
    \item $A_{ij} = \left[0 \ldots  -\Delta p_{ij}^\top \ldots \Delta p_{ij}^\top \ldots  0\right]$ with the corresponding states of agents $i$ and $j$ inserted in the matrix,\\
    \item $b_{ij} = \gamma h_{ij}^3 + \frac{\Delta v_{ij}^\top \Delta p_{ij}}{2(\alpha_i + \alpha_j)} - \frac{(\Delta v_{ij}^\top \Delta p_{ij})^2}{2 \|\Delta p_{ij}\|^2} + \frac{\|\Delta v_{ij}\|^2}{2}$,\\
    \item $\gamma > 0$ is a controller parameter that regulates the approach to the safety boundary.
\end{itemize}
These constraints ensure that agent $i$’s control input $u_i$ and agent $j$’s control input $u_j$ avoid collisions by satisfying the safety barrier certificates, integrated within a quadratic programming (QP) framework for real-time applications.

Note: In the definition of $b_{ij}$, $\gamma$, $h_{ij}$, $\Delta p_{ij}$, and $\Delta v_{ij}$ correspond to the controller parameter, the CBF function for agents $i$ and $j$, relative position vector, and the relative velocity vectors among the two agents, respectively. The symbols $\alpha_i$ and $\alpha_j$ represent the acceleration constraints of agents $i$ and $j$.
\subsection{Deadlock Condition}
\label{subsec:deadlock_condition}

\begin{figure}[htbp]
    \centering
    \includegraphics[width=0.4\textwidth, height=0.25\textwidth]{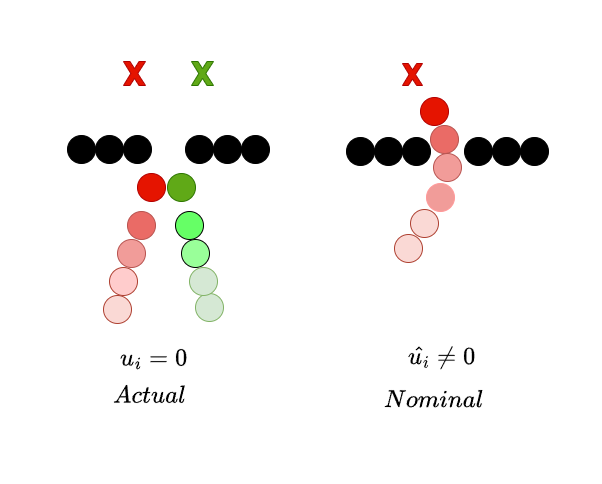}
    \caption{Condition for deadlock among agents to occur}
    \label{fig:deadlock}
\end{figure}

Deadlock scenario in multi-agent navigation systems primarily arise due to the formation of symmetries in the environment that may lead to conflicts between agents. Such symmetries often manifest in scenarios where multiple agents, following identical or mirror-like navigation policies, encounter each other in narrow passages, intersections, or areas with restricted maneuverability. The mathematical expression for identifying a deadlock situation hinges on the actual control inputs \(u_i\) and the nominal control inputs \(\hat{u}_i\). The nominal control inputs represent the intended direction and magnitude of movement in the absence of other agents in the given scenario. In contrast, the actual control inputs are adjusted based on the real-time navigation strategy, incorporating obstacle avoidance and agent-to-agent spacing requirements. Fig.~\ref{fig:deadlock} depicts the actual and nominal scenario along with the behaviour of controls which enforces deadlock to happen, equation~\ref{eq:deadlock_condition} shows the condition for the deadlock to happen.

\begin{equation}
    \text{Deadlock Condition:} \quad u_i = 0 \quad \text{and} \quad \hat{u}_i \neq 0
\label{eq:deadlock_condition}
\end{equation}

\section{Proposed Methodology}
We introduce the Strategic Pseudo-Goal Perturbation (SPGP) framework, which is specifically designed to resolve deadlock in a centralized multi-agent environment through the use of pseudo goals. SPGP synthesizes strategic navigation with robust safety mechanisms to negotiate complex social interactions between agents.

\subsection{Mathematical Formulation}
\label{subsec:mathematical_formulation}
In the Strategic Pseudo-Goal Perturbation (SPGP) framework, we analyse the problem with efficiently navigating a given set of autonomous agents \( \mathcal{A} = \{1, 2, \ldots, N\} \) towards their respective targets within a given environment. Each agent \( i \), starting from an initial position \( \mathbf{p}_i(t) \), is tasked with reaching a designated goal \( \mathbf{g}_i \). The SPGP framework aims to optimize the planned path so as to ensuring deadlock and collision avoidance and generating optimal path for the agents. This is achieved by generating the agent's trajectories using Safety Barrier Certificates (SBCs), which enforces a safety certificates using CBFs around each agent to prevent inter-agent collisions.

\subsubsection{Safety Barrier Certificates}
\label{subsubsec:safety_barrier_certificates}
The SBCs are important in the SPGP's collision avoidance strategy, imposing constraints that dynamically adapt to the changing states and control of the agents. Given any two agents \( i \) and \( j \), the safety constraints are formulated in the equation ~\ref{eq:sbc}:
\begin{equation}
    h_{ij}(t) = \| \mathbf{p}_i(t) - \mathbf{p}_j(t) \| - (r_i + r_j) \geq 0,
\label{eq:sbc}
\end{equation}
where \( h_{ij}(t) \) denotes the euclidean distance between the two agents, and \( r_i \) and \( r_j \) represents their respective safety radius. This function ensures that the agents maintain a separation distance which should be greater than the sum of their respective safety radius, effectively establishing a collision-free path for the agents. 

\subsection{Pseudo-Goal Selection Strategy}
\label{subsubsec:pseudo_goal_selection}
When the agents actual controls are below a certain threshold ( \( \mathbf{u}_t \)) then the deadlock occurs in the simulation. When deadlock occurs the pseudo-goal selection strategy plays an instrumental role in alleviating deadlock conditions that materialize when paths of agents intersect, leading to potential immobilization. This strategy ingeniously circumvents such predicaments by allocating an interim waypoint, designated as a pseudo-goal (\( \mathbf{pg}_i \)), towards which an agent can momentarily divert its course.

\begin{figure}[htbp]
    \centering
    \includegraphics[width=0.35\textwidth, height=0.35 \textwidth]{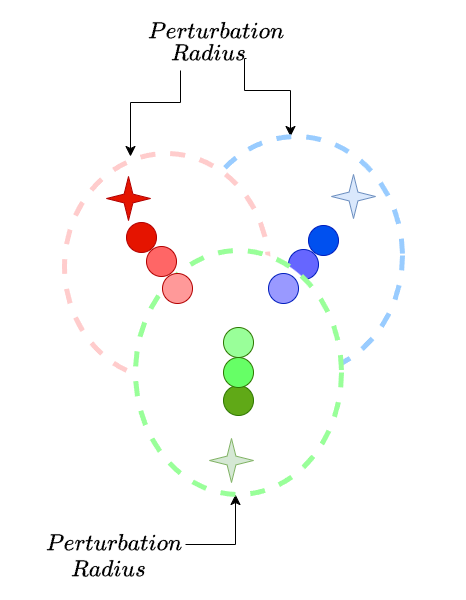}
    \caption{Pseudo goal selection to resolve deadlock and enforce separation between agents}
    \label{fig:perturb}
\end{figure}

The framework develops the navigable region \( \mathcal{B}(\mathbf{p}_i, \delta) \) as a circular section, the centre is positioned on the agent's existing position, with radius \( \delta \) defining its extent. Within this area, a pseudo-goal is selected based on generating a random point \(x ,y\) that aims to maximize the distance from surrounding agents, in alignment with the safety objectives for barrier certificates . The given procedure can be mathematically represented by the equation~\ref{eq:pg}:
\begin{equation}
    \mathbf{pg}_i = \mathbf{p}_i + \delta \times \begin{bmatrix} \cos(2\pi \times \text{x}) \\ \sin(2\pi \times \text{y}) \end{bmatrix},
\label{eq:pg}
\end{equation}
Fig.~\ref{fig:perturb} depicts the strategic pseudo goal selection in a circle to resolve deadlock among the agents and enforce the maximum separation between agents.

Upon the identification of a feasible pseudo-goal, the agent updates its control inputs to navigate towards this new temporary target.The resulting trajectory is then a smooth, deadlock-free path that respects the dynamic constraints of the agent and is robust to uncertainties in the environment. As agents move towards their pseudo-goals, the path adjustment mechanism continuously updates the planned path to resolve the changing states of other agents and potential new obstacles in this process. 

After reaching the pseudo-goal, agents resets their navigation systems back towards the primary target location. Algorithm~\ref{alg:RP} summarizes the Strategic Pseudo-Goal Perturbation (SPGP) algorithm and it's working to resolve the possible deadlocks.

\subsection{Optimization Problem Formulation}
The mathematical formulation of this optimization problem uses the core of Quadratic Programming (QP) expressed in equation~\ref{eq:optimized_control_input}. This approach tries to minimize the cost function \( C(\mathbf{u}) \), which serves as the main objective of our optimization. The objective function, represented in the equation ~\ref{eq:optimized_control_input}, aims to reduce the squared deviation for the actual control inputs, \( \mathbf{u}_i \), out of the nominal scenario, \( \hat{\mathbf{u}}_i \). This minimization of cost function leads in achieving the scenario to mirror the nominal controller's trajectory, ensuring that each state remains as close to the intended path as feasibly possible, thus maximizing the accuracy and efficiency.

\begin{equation}
\begin{aligned}
    \mathbf{u}^* &= \underset{\mathbf{u} \in \mathbb{R}^{2N}}{\text{argmin}} \; C(\mathbf{u}) = \sum_{i=1}^{N} \| \mathbf{u}_i - \hat{\mathbf{u}}_i \|^2 \\
    &\text{s.t.} \\
    &\quad A_{ij}\mathbf{u} \leq b_{ij}, \quad \forall i \neq j, \\
    &\quad \| \mathbf{u}_i \|_2 \leq \alpha_i, \quad \forall i \in \mathcal{A}.
\end{aligned}
\label{eq:optimized_control_input}
\end{equation}

This formulation considers the necessary constraints which includes \( A_{ij}\mathbf{u} \leq b_{ij} \) condition that represents a safety barrier, which makes sure that each agent maintains a safe distance from its neighbouring agents, thereby reducing potential collisions. Furthermore, the condition \( \| \mathbf{u}_i \|_2 \leq \alpha_i \) imposes a maximum limit on the magnitude of control inputs, confining all the agents within the practical and feasible values of dynamic capabilities .

\begin{algorithm}
\caption{Strategic Pseudo-Goal Perturbation (SPGP)}
\begin{algorithmic}[1]
\State Initialize  \( \mathbf{p}_i(t) \),  \( \mathbf{g}_i \), \( \mathbf{u}_t \)
\State $u_i \gets$ actual controls
\State $\hat{u_i} \gets$ nominal controls
\State Deadlock\_Flag $\gets$ False
\State temp\_var $\gets$ \( \mathbf{g}_i \)   
\State Run SBC
\If{$u_i \leq u_t$ \textbf{and} $\hat{u_i} > u_t$}   
    \State Deadlock\_Flag $\gets$ True
\EndIf
\If{Deadlock\_Flag = True}
    \State  \( \mathbf{g}_i \) $\gets$\( \mathbf{pg}_i \)
\EndIf
\State Update \( \mathbf{p}_i(t) \)
\State $x, y \gets$ \( \mathbf{pg}_i \)
\State \( \mathbf{g}_i \) $\gets$ temp\_var
\State Run SBC Again
\end{algorithmic}
\label{alg:RP}
\end{algorithm}

\section{Results and Discussions}

This section shows outcomes and comparative study of our Strategic Pseudo-Goal Perturbation (SPGP) framework against a suite of established methodologies within the multi-agent navigation domain. Central to our evaluation is the examination of Success rate (SR), average velocity change ($\Delta$V) and path deviation across four challenging social mini-games scenarios. The optimization of agent trajectories, facilitated by Quadratic Programming (QP), seeks to minimize deviations from nominal controls, effectively mirroring the nominal controller's behavior.

\begin{figure}[htbp]
    \centering
    \begin{subfigure}[b]{0.48\linewidth}
        \includegraphics[width=\linewidth]{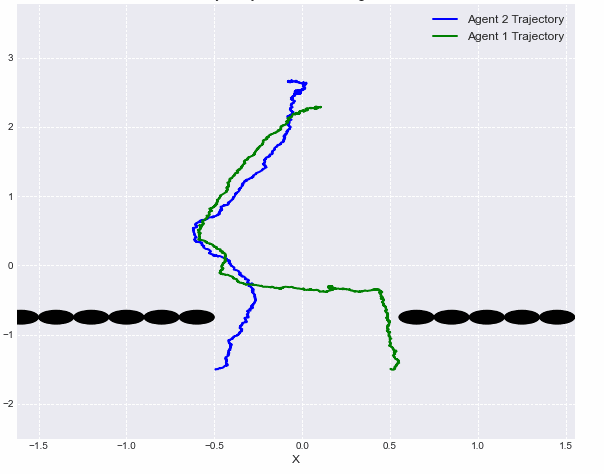}
        \caption{Doorway setting} 
    \end{subfigure}
    \hfill 
    \begin{subfigure}[b]{0.48\linewidth}
        \includegraphics[width=\linewidth]{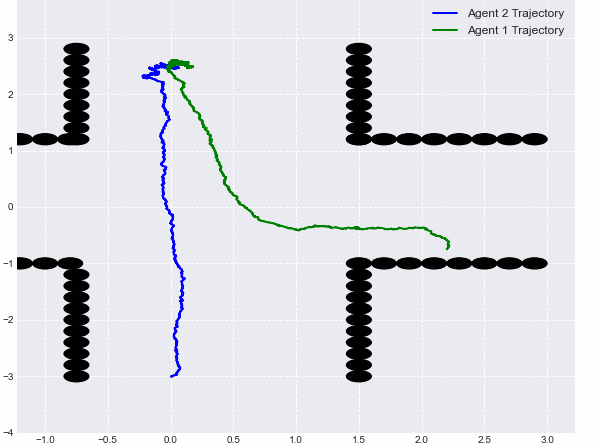}
        \caption{Intersection setting} 
    \end{subfigure}

    \begin{subfigure}[b]{0.48\linewidth}
        \includegraphics[width=\linewidth]{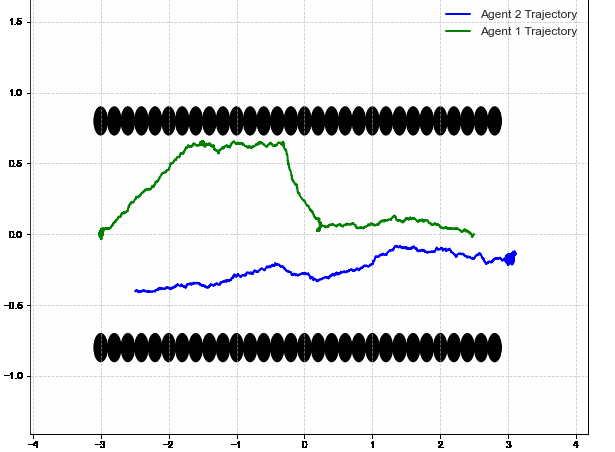}
        \caption{Hallway setting}
    \end{subfigure}
    \hfill 
    \begin{subfigure}[b]{0.48\linewidth}
        \includegraphics[width=\linewidth]{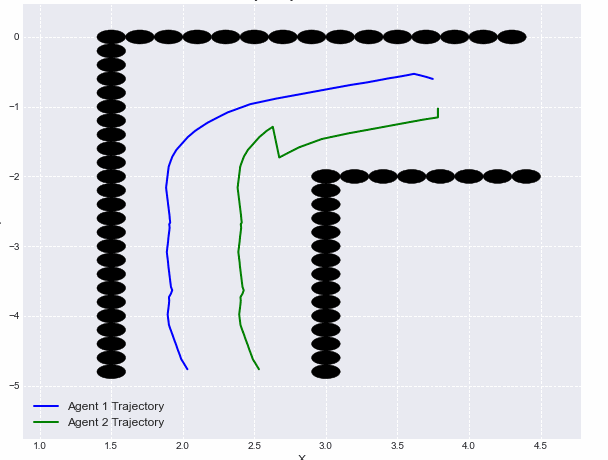}
        \caption{L corner setting} 
    \end{subfigure}
    
    \caption{Planned trajectories of two agent in social navigation settings using strategic perturbation}
    \label{fig:scenarios}
\end{figure}

\begin{table}[!htbp]
\caption{Performance Comparison Across Different Navigation Strategies}
\tiny
\centering
\begin{tabular}{|c|c|c|c|c|}
\hline
\textbf{Scenario} & \textbf{Method} & \textbf{SR(\%)} & \textbf{Avg. $\Delta$V} & \textbf{Path Deviation} \\
\hline
\multirow{4}{*}{Doorway} & SBC & $0.00 \pm 0.00$ & $0.10 \pm 0.01$ & $0.16\pm 0.02$ \\
& ORCA & $50.00 \pm 0.00$ & $0.100 \pm 0.00$ & $0.000\pm 0.00$ \\
& NH-TTC & $0.00 \pm 0.00$ & $0.2 \pm 0.00$ & $0.160\pm 0.00$ \\
\rowcolor{gray!25}& \textbf{SPGP (Ours)} & $\mathbf{100 \pm 0.00 }$ & $\mathbf{0.03 \pm 0.02}$ & $\mathbf{0.187\pm 0.02}$ \\
\hline
\multirow{4}{*}{Intersection} & SBC & $0.00 \pm 0.00$ & $0.19\pm 0.01$ & $0.08\pm 0.005$ \\
& ORCA & $50.00 \pm 0.00$ & $0.250 \pm 0.00$ & $0.89 \pm 0.00$ \\
& NH-TTC & $50.00 \pm 0.00$ & $0.02 \pm 0.00$ & $0.47\pm 0.00$ \\
\rowcolor{gray!25}&\textbf{SPGP (Ours)} & $\mathbf{100 \pm 0.00 }$ & $\mathbf{0.034 \pm 0.01}$ & $\mathbf{0.4 \pm 0.14}$ \\
\hline
\multirow{4}{*}{Hallway} & SBC & $0.00 \pm 0.00$ & $0.005 \pm 0.003$ & $0.47\pm 0.05$ \\
& ORCA  & $100.00 \pm 0.00$ & $0.110 \pm 0.00$ & $1.990\pm 0.00$ \\
& NH-TTC & $100 .00 \pm 0.00$ & $0.02 \pm 0.00$ & $0.35 \pm 0.00 $ \\
\rowcolor{gray!25}&\textbf{SPGP (Ours)} & $\mathbf{100 \pm 0.00 }$ & $\mathbf{0.05 \pm 0.01}$ & $\mathbf{0.32\pm 0.32}$ \\
\hline
\multirow{4}{*}{L corner} & SBC & $25.00 \pm 25.00$ & $0.11 \pm 0.01$ & $0.65 \pm 0.02$ \\
& ORCA & $50.00 \pm 0.00$ & $0.28 \pm 0.00$ & $0.99 \pm 0.00$ \\
& NH-TTC & $100 .00 \pm 0.00$ & $0.02 \pm 0.00$ & $0.32 \pm 0.00 $ \\
\rowcolor{gray!25}&\textbf{SPGP (Ours)} & $\mathbf{100 \pm 0.00 }$ & $\mathbf{0.026 \pm 0.02}$ & $\mathbf{0.69\pm 0.35}$ \\
\hline
\end{tabular}
\label{tab:performance_comparison}
\end{table}

Table ~\ref{tab:performance_comparison} shows the results for the SPGP and other baseline methods for Deadlock Resolution, Average change in velocity and Path difference. Due to random nature of our algorithm we carried out each scenario experiment 10 times and recorded the mean values and standard deviation values of the scores. The perturbation radius is taken as 1m and safety radius is taken as 0.2m for all agents.  Our approach is meticulously benchmarked against counterparts such as SBC, ORCA and NH-TTC,  Success rate measures the efficacy of reaching the goal position without and deadlock and collision, average change in velocity for each scenario is shown which should be minimal while navigation and path deviation is the difference in trajectory of actual and nominal case for the concerned scenario. It is measured by calculating the Hausdorff distance between actual and nominal trajectory. Notably, the SPGP framework depicts higher performance in achieving success as compared with different other methods, further SPGP also yields positive results in minimizing velocity changes and path deviations, thereby asserting its efficacy in ensuring smooth and efficient agent navigation. The trajectory for two agent scenario is depicted in Fig.~\ref{fig:scenarios}. For doorway and intersection scenarios, the goal position for both agents were same. The SPGP successfully reaches the specified goal position without any collision and deadlock by enforcing the perturbation strategy. By ablation study, we concluded that SPGP is easily scalable upto 8 agents in the doorway setting, 10 agents in the intersection and hallway setting, and 5 agents in the L-corner setting however it was noted that on increasing the number of agents the time to goal for each agent increases which increases the simulation time and the maximum number of agent that each setting can handle vastly depends on the geometric properties of the social min-game scenarios.

\begin{figure}[htbp]
    \centering
    \includegraphics[width=0.45\textwidth, height=0.25\textwidth]{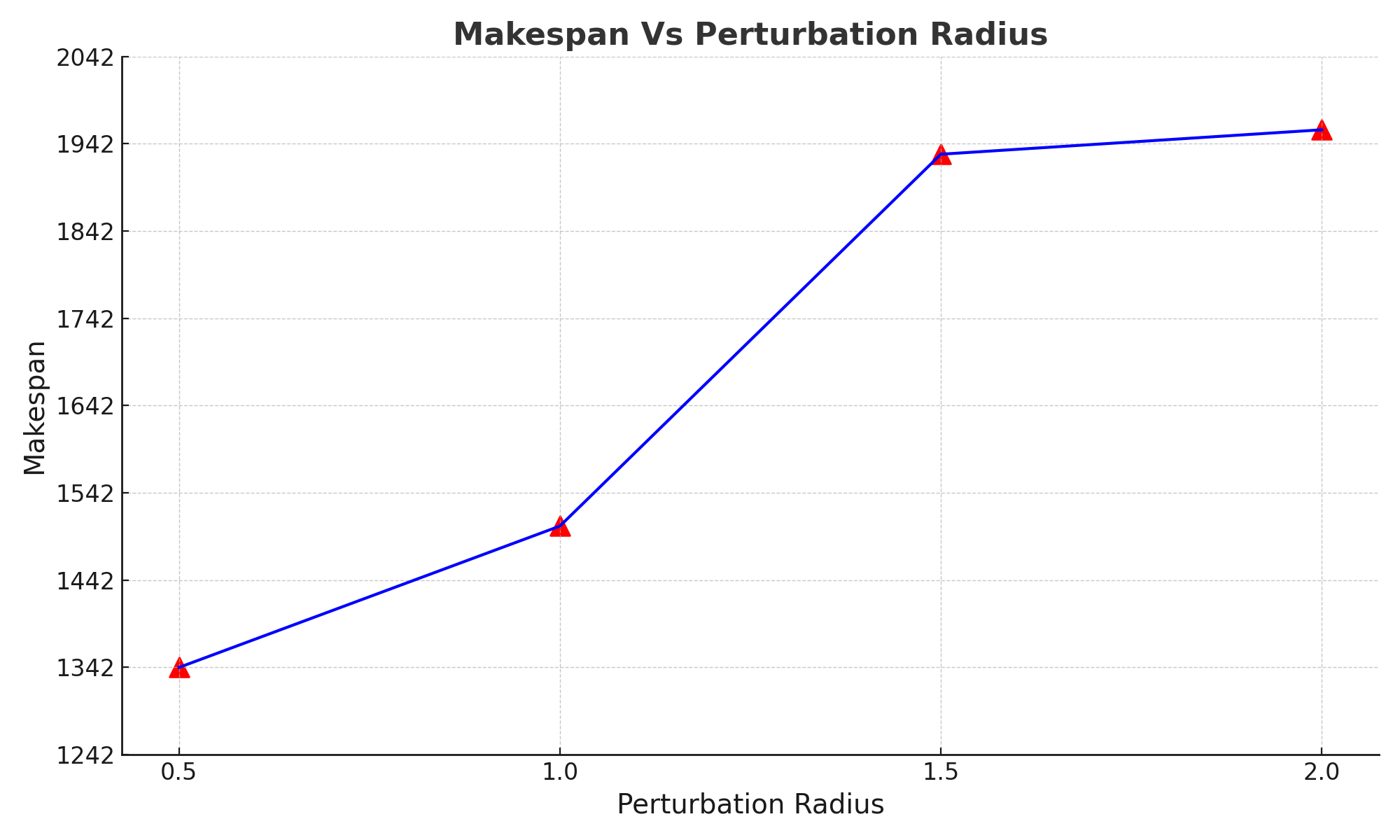}
    \caption{Effect of different perturbation radius on the makespan of agent.}
    \label{fig:makespan}
\end{figure}

Another experiment is performed to analyse the the effect on performance with variation of perturbation radius. In SPGP based on the perturbation radius the agent's perturbs from it's designated path to avoid deadlock and this effects the overall time in which the given agent's arrives at the goal position. To analyse this we measured the variation of makespan  with different perturbation radius. The makespan value shows the time or timesteps required for the given agent to arrive at the goal position.

Fig~\ref{fig:makespan} shows the change in makespan as the perturbation radius is varied. It is observed that on decreasing the perturbation radius the makepsan decreases, this happens because the agent's pseudo goal selection is confined to a circle of small radius. Also it was observed that after perturbation radius of 1.5m their is not much increase in makespan indicating that agent is not selecting too far away pseudo goal position so as to enforce distance between the deadlocked agents.

\section{Conclusion and Future Work}
The study introduced a unique Strategic Pseudo-Goal Perturbation (SPGP) method, integrating Safety Barrier Certificates (SBCs) and a strategic pseudo-goal strategy to improve multi-agent navigation efficiency in complex social scenarios. Our findings, validated through diverse simulation scenarios, underscore SPGP's effectiveness in reducing makespan and enhancing path optimization. Future endeavors will focus on incorporating machine learning for predictive congestion management, extending the framework to heterogeneous agent systems for advanced coordination, and empirical validation in real-world settings to improve its application. Another extension of this work can be to integrate the SPGP framework with multi-agent reinforcement learning approaches and evaluate it on different social mini-game scenarios to check the method's efficacy for deadlock avoidance.

\end{document}